\documentstyle[12pt]{article}

\newcommand{\eq}{\begin{equation}}
\newcommand{\en}{\end{equation}}
\newcommand{\eqn}{\begin{eqnarray}}
\newcommand{\enn}{\end{eqnarray}}

\newcommand{\beq}{\begin{equation}}
\newcommand{\eeq}{\end{equation}}

\begin{document}

\begin{titlepage}

\begin{flushright}
  hep-th/9906223\\
USC-99/HEP-B2 \\
\end{flushright}
\vspace{2cm}
\begin{center}

{\bf STRINGS, BRANES and TWO-TIME PHYSICS \footnote{
This research was partially supported by the U. S. Department of
 Energy under grant number DE-FG03-84ER40168 and by a National 
Science Foundation grant number NSF9724831 for collaborative
 research between USC and Japan. } } \\
\vspace{1cm}
{\bf I. Bars, C. Deliduman and D. Minic\footnote{e-mail:
bars@physics.usc.edu,
 cdelidum@physics.usc.edu, minic@physics.usc.edu} } \\
\vspace{.5cm}
Department of Physics and Astronomy\\
University of Southern California\\
Los Angeles, CA 90089-0484 \\
\vspace{1cm}
{\bf Abstract}

\end{center}

We generalize the ideas and formalism of Two-Time Physics from particle
dynamics to some specific examples of string and p-brane
($ p \ge 1 $) dynamics. The two-time string or p-brane action can be gauge
fixed to produce various one-time string or p-brane actions that are dual
to each other under gauge transformations. 
We discuss the particular gauges that correspond to tensionless
strings and p-branes in flat (d-1)+1 spacetime, rigid strings and p-branes 
 in flat (d-1)+1 spacetime, and tensionless strings
and p-branes propagating in the AdS$_{d-n} \times $ S$^n $ backgrounds.

\end{titlepage}

\section{Introduction}

There exist by now many convincing examples of a remarkable connection
between a global $SO(d,2)$ symmetry (sometimes hidden and non-lineraly
realized) in one-time point particle dynamics and the two-time Lorentz SO$%
\left( d,2\right) $ symmetry that is linearly realized in a two-time action
with extra gauge symmetries \cite{bda}-\cite{bdm1}. The basic property is
that the gauge symmetry reduces the two-time action to a one-time action,
but in many possible gauge equivalent ways\footnote{%
Note that, unlike Yang-Mills type gauge theories, the concept of ``time'' is
gauge dependent in the Two-Time-Physics formalism (as in e.g. General
Relativity). Therefore the gauge fixed Lagrangian or Hamiltonian describes
diverse systems that would appear unrelated in ordinary 1-time dynamics.
However, there are gauge invariant quantities that are the same in every one
of these seemingly unrelated systems, thus displaying their unified origin.}%
, thus establishing a ``duality'' between various one-time dynamical systems
and unifying them in some sense. The motivation for this approach - called
Two-Time Physics - comes from the algebraic approach to M-theory and its
extensions \cite{bars1, mth, fth, bk, bd2} that provide various hints of
hidden two timelike dimensions. There also exist indications that the
formalism of Two-Time Physics is applicable to the full structure of
M-theory \cite{bdm2}.

The crucial observation of \cite{bda}-\cite{bdm1} is that many point
particle systems (relativistic massless and massive particle,
non-relativistic massless and massive particle, H-atom, harmonic oscillator,
particle moving in the AdS background, etc.) in (d-1)-space and 1-time
dimensions can be understood as different gauge choices of a 2-time point
particle system endowed with a local $Sp(2,R)$ symmetry and manifestly
invariant under the global $SO(d,2)$ symmetry. Unlike Yang-Mills type gauge
theories, the concept of ``time'' is gauge dependent in the Two-Time-Physics
formalism (see also General Relativity although it has only one gauge
dependent timelike dimension). Therefore the gauge fixed Lagrangian or
Hamiltonian (spectrum) describes diverse systems that would appear unrelated
in ordinary 1-time dynamics. However, there are gauge invariant quantities
that can be computed in any one of these systems with identical results (for
example Casimir eigenvalues of the SO$\left( d,2\right) $ representation).
Thus such systems display properties of their unification under a common
gauge invariant action.

The particular $Sp(2,R)$ symmetry that is gauged in this case is nothing but
the familiar authomorphism of the Heisenberg commutation relations in
quantum mechanics, or the symplectic group of the one-particle phase space
of classical Hamiltonian dynamics. One can also think of the same $%
Sp(2,R)\sim SO(1,2)$ as the conformal group in one (world-line) dimension.
From this point of view many $0+1$ quantum gravity systems (relativistic
massless and massive particle, non-relativistic massless and massive
particle, etc.) can be viewed as different gauge choices of the same $0+1$ 
{\it conformal} quantum gravity theory \cite{bda}.

In some specific instances (like the relativistic massless particle) the
action of the global $SO(d,2)$ can be interpreted as the natural action of
the conformal group in d-dimensions. What is surprising is that the same
group has a natural action, even though with a highly non-linear
realization, in many other physical situations (the relativistic massive
particle, for example) \cite{b1}. The Hamiltonians describing these two
different physical systems (relativistic massless and massive particle) are
related by a canonical $Sp(2,R)$ transformation, which is responsible for
the appearance of the global $SO(d,2)$ invariance of the action describing
the relativistic massive particle. The evolution parameters (``times'') in
the two cases, being canonically conjugate to the respective Hamiltonians,
are different as well.

One can also interpret the same $SO(d,2)$ as the Lorentz group in the space
with one extra spatial and one extra time dimensions. The parameters in the
various Hamiltonians related by the local $Sp(2,R)$ transformations, appear
as moduli (values of fields) in this $(d,2)$ dimensional space (e.g. mass of
particle, coupling of H-atom, etc.). Note that the Two-Time Physics
formalism works both classically and quantum-mechanically \cite{bda}-\cite
{bdm1}. It turns out that in the full quantum theory, the $Sp(2,R)$ gauge
invariant sector is completely described by a unique representation of the
conformal group $SO(d,2)$, which fixes the values of the $SO(d,2)$ Casimirs.

The same formalism can be generalized to the case of particles with spin $n/2
$ by using world-line supersymmetry with $n$ supercharges (by replacing $%
Sp(2,R)$ with $OSp(n|2)$) \cite{bd1} and to target-space supersymmetry (by
generalizing the action of the global $SO(d,2)$ symmetry to various
superconformal groups) \cite{bdm1}. In the latter case, the formalism of
Two-Time Physics sheds new light on the origin of fermionic kappa
supersymmetry. This provides an explicit way for constructing physical
models invariant under global superconformal symmetry groups OSp$\left(
8/4^{\ast }\right) $, SU$\left( 2,2/4\right) $, OSp$\left( 8^{\ast
}/4\right) $ which have usual space-time interpretations in $d=3,4,6$ \cite
{bdm1} and an expanded space-time interpretation including zero modes of
p-brane gauge potentials in other dimensions \cite{bdm2}. The same approach
explicitly establishes the existence of the novel bosonic counterparts of
kappa supersymmetry \cite{bdm2}.

In this letter we want to extend the formalism of Two-Time Physics for the
case of strings and p-branes ($p\ge 2$). We find a natural action of the
global $SO(d,2)$ conformal symmetry in the case of many classical systems
involving bosonic strings and branes in $\left( d-1\right) +1$ dimensions.
The list includes the bosonic null string and branes, the null string moving
in the AdS background and the bosonic rigid string and branes. The actions
for these systems admit the extension to the two-time formalism by the
addition of gauge degrees of freedom. The new gauge invariant actions can be
gauge fixed to many other dual sytems of branes in the same manner of
particle dynamics.

The letter is organized as follows. In section 2 we briefly review the
Two-time Physics formalism for the case of particle dynamics. Then in
section 3 we extend this formalism and discuss tensionless strings and
branes by working with configuration space variables only. We apply this
approach in section 4 for the case of tensionless strings and branes moving
in the $AdS$ background. In section 5 we discuss the rigid bosonic strings
and branes. Finally, we comment on possible generalizations of the results
presented in this letter.

\section{Basic Formalism of Two-Time Physics}

We start with a brief summary of the Two-Time Physics formalism for the case
of spinless particle dynamics \cite{bda}. The model is an $Sp(2,R)$ gauge
theory described by the action (the various forms of the action and their
utility will be explained below) 
\begin{eqnarray}
S_0 &=&\frac 12\int d\tau \,D_\tau X_i^MX_j^N\varepsilon ^{ij}\eta _{MN}
\label{action} \\
&\equiv &\int d\tau (\partial _\tau X_1^MX_2^N-\frac 12A^{ij}X_i^MX_j^N)\eta
_{MN}  \label{second} \\
&=&\int d\tau \left[ \partial _\tau X^MP^M-\frac 12eP^MP^M-AX^MP^M-\frac
12kX^MX^M\right] \eta _{MN} \\
&\equiv &\int d\tau \left[ \frac 1{2e}\left( \partial _\tau X-AX\right)
^2-\frac 12k\,X^2\right]  \label{conform} \\
&\equiv &\int d\tau \left[ \frac 1{2e}\left( \partial _\tau X\right)
^2-\frac 12K\,X^2\right] ,\quad K=k-\frac{A^2}e-\partial _\tau \left( \frac
Ae\right)  \label{confor}
\end{eqnarray}
Here $X_i^M(\tau )$ is an $Sp(2,R)$ doublet, consisting of the ordinary
coordinate and its conjugate momentum ($X_1^M\equiv X^M$ and $X_2^M\equiv
P^M $). The indices $i,j=1,2$ denote the doublet $Sp(2,R)$, they are raised
and lowered by the antisymmetric Levi-Civita symbol $\varepsilon _{ij}$. The
covariant derivative $D_\tau $ is defined as 
\begin{equation}
D_\tau X_i^M=\partial _\tau X_i^M-\varepsilon _{ik}A^{kl}X_l^M.
\end{equation}
The local $Sp(2,R)$ acts as $\delta X_i^M=\varepsilon _{ik}\omega ^{kl}X_l^m$
and $\delta A^{ij}=\omega ^{ik}\varepsilon _{kl}A^{kj}+\omega
^{jk}\varepsilon _{kl}A^{ik}$, where $\omega ^{ij}\left( \tau \right) $ is a
symmetric matrix containting three gauge parameters. The second form of the
action is obtained after an integration by parts so that only $X_1^M$
appears with derivatives. This allows the identification of $X,P$ by the
canonical procedure, as indicated in the third form of the action.

The gauge fields $A^{11}\equiv k$, $A^{12}=A^{21}\equiv A$, and $%
A^{22}\equiv e$ also act as Langrange multipliers for the following three
first class constraints (that form the Sp$\left( 2,R\right) $ algebra) 
\begin{equation}
X_i\cdot X_j=0\,\,\rightarrow \,\,\,X^2=P^2=X\cdot P=0,
\end{equation}
as implied by the local $Sp(2,R)$ invariance. It is precisely the solution
of these constraints that require that the global metric $\eta _{MN}$ has a
signature with two-time like dimensions. Thus, $\eta _{MN}$ stands for the
flat metric on a ($d,2$) dimensional space-time, which is the only signature
consistent with the equations of motion for the $Sp(2,R)$ gauge field $%
A^{kl} $, leading to a non-trivial dynamics that can be consistently
quantized. Hence the global two-time $SO(d,2)$ is implied by the local $%
Sp(2,R)$ symmetry. It is possible to extend $\eta _{MN}$ to curved spacetime 
$G^{MN}\left( X\right) $ with suitable modification of the transformation
laws and certain conditions on the metric. Likewise, it is possible to
extend the action to include background gauge fields. Such generalizations
will be discussed in \cite{emgrav}.

The fourth form of the action (\ref{conform}) is obtained by integrating out 
$P^M$. Note that the ordinary $\tau $ reparametrization $\tau \rightarrow
\tau +\delta \tau \left( \tau \right) $ corresponds to a $U(1)$ subgroup of $%
Sp(2,R)$, with the gauge potential $A^{22}=e$ and gauge parameter $\omega
^{22}=e\left( \tau \right) \delta \tau \left( \tau \right) $. Recalling that
Sp$\left( 2,R\right) =SO\left( 1,2\right) $ is the conformal group on the
worldline, this form of the action can be viewed as a generalization of $0+1$
gravity ($\tau $-reparametrization) to $0+1$ {\it conformal} gravity. The
gauge potentials $e,A,k$ and parameters $\omega ^{22},\omega ^{12},\omega
^{11}$ are associated with the SO$\left( 1,2\right) $ generators for local
translations, dilatations and conformal transformations respectively.

The fifth form of the action (\ref{confor}) is obtained after integtating by
parts the cross term $-\frac AeX\cdot \partial _\tau X$ and defining $K=k-%
\frac{A^2}e-\partial _\tau \left( \frac Ae\right) $ as the total coefficient
of $X^2/2$. This form of the action is still invariant under the same Sp$%
\left( 2,R\right) =SO\left( 1,2\right) $ local symmetries. The simplest form
of the symmetries are the $\tau $ reparametrization $\delta \tau \left( \tau
\right) $ and the local scaling $\omega ^{12}\equiv \alpha \left( \tau
\right) $%
\begin{eqnarray}
\delta X^M &=&\delta \tau \partial _\tau X^M,\quad \delta K=\partial _\tau
\left( K\,\delta \tau \right)  \label{dtau} \\
\delta _\alpha X^M &=&\alpha X^M,\quad \delta _\alpha e=2\alpha \,e,\quad
\delta _\alpha K=-2\alpha \,K+\partial _\tau \left( \frac{\partial _\tau
\alpha }e\right) ,  \label{dalpha}
\end{eqnarray}
while $X^M,e$ and $K$ are all invariant under the local conformal
transformations $\omega ^{11}$. This is the simplest form of the action, and
it is this form that we will explore below in discussing strings and
p-branes.

All forms of the action (\ref{action}-\ref{confor}) have an explicit global $%
SO(d,2)$ invariance, generated by the Lorentz generators $%
L^{MN}=X^MP^N-X^NP^M=\varepsilon ^{ij}X_i^MX_j^N$ that are gauge invariant.
As we mentioned above, different gauge choices lead to different $0+1$
theories of $0+1$ gravity (the relativistic massless and massive particles,
H-atom, harmonic oscillator, etc.) all of which have $SO(d,2)$ invariant
actions that are directly obtained from (\ref{action}-\ref{confor}) by gauge
fixing. Since the action (\ref{action}-\ref{confor}) and the generators $%
L^{MN}$ are gauge invariant, the global symmetry SO$\left( d,2\right) $ is
not lost by gauge fixing. This explains why one should expect a hidden
(previously unnoticed) global symmetry SO$\left( d,2\right) $ for each of
the $0+1$ systems that result by gauge fixing. Note that the equations of
motion, not only the action, of the manifestly $SO(d,2)$ invariant particle
system reduce to the expected equations of motion of the $(d-1,1)$ gauge
fixed systems, indicating the validity of the gauge-fixing procedure.

If the system is quantized in a fixed gauge the absence of ghosts is
evident. The system can also be quantized covariantly; then the $Sp(2,R)$
gauge symmetry is just enough to remove all negative-norm states
(``ghosts'') introduced by two timelike dimensions so that the resulting
quantum theory is unitary. In the quantum version of the theory the unifying
feature of the system is displayed by the fact that all the gauge fixed
systems, or the covariantly quantized system, share the same unitary
representation of SO$\left( d,2\right) $. For example the quadratic Casimir
eigenvalue is $C_2\left( SO\left( d,2\right) \right) =1-d^2/4$ for the free
massless particle, the H-atom, harmonic oscillator (in one less dimension),
the particle moving on AdS$_{d-n}\times S^n$ (any $n$), etc., and similarly
for all Casimir eigenvalues $C_n\left( SO\left( d,2\right) \right) $.

As mentioned before this action has been generalized to particles with spin $%
n/2$ (via local OSp$\left( n/2\right) $ \cite{bd1}), to spacetime
supersymmetry \cite{bdm1} \cite{bdm2} and to interactions with background
fields \cite{emgrav}.

\section{Extensions to Tensionless Strings and Branes}

In this section we want to extend the formalism of Two-Time Physics reviewed
in the previous section for the case of tensionless strings and branes. We
do not start from a general formalism as in the case of particle mechanics
(the general formalism which is still being developed will be given
elsewhere). Rather, we discuss specific cases, which point toward a more
general formulation. We will work with configuration space variables as in
the the last form of the action (\ref{confor}).

We begin with a string analog of the massless relativistic particle. We
rotate the extra space $X^{1^{\prime }}$ and time $X^{0^{\prime }}$
dimension into a pair of ``light-cone-like'' coordinates $X^{+^{\prime }}$
and $X^{-^{\prime }}$.We recall that \cite{bda}-\cite{bdm1} in this basis $%
M=\left( +^{\prime },-^{\prime },\mu \right) $, fixing two gauges $%
X^{+^{\prime }}\left( \tau \right) =1$, $P^{+^{\prime }}\left( \tau \right)
=0$, and solving two constraints $X^{2}=X\cdot P=0$, gives $X^{-^{\prime
}}=x^{2}/2$ after identifying $X^{\mu }=x^{\mu }.$ Then the action (\ref
{action}-\ref{confor}) reduces to the action for a massless particle in $d$%
-dimensions $S=\int d\tau {\frac{{\dot{x}}_{\mu }^{2}}{{2e}}}$. This action
has a well known global non-linearly realized conformal symmetry SO$\left(
d,2\right) $ which is a reflection of the SO$\left( d,2\right) $ two-time
Lorentz symmetry of the original gauge invariant action as as explained in
the previous section. This action implies the familiar equations of motion $%
\ddot{x}_{\mu }=0$ and $\dot{x}^{2}=0$. The stringy counterpart of the
massless particle is the null string \cite{null} in $d$-space-time
dimensions 
\begin{equation}
S_{ns}=\int d^{2}\sigma {\frac{{\{x^{\mu },x^{\nu }\}}^{2}}{{2e}},}
\label{nulls}
\end{equation}
where we used the Poisson bracket $\{x^{\mu },x^{\nu }\}=\epsilon
^{ab}\partial _{a}x^{\mu }\partial _{b}x^{\nu }$ with $a$ or $b=0,1$
denoting the world-sheet indices for $\tau $ and $\sigma $. The equations of
motion for $x^{\nu }\left( \tau ,\sigma \right) $ and $e\left( \tau ,\sigma
\right) $ are respectively 
\[
\{x_{\mu },\frac{\{x^{\mu },x^{\nu }\}}{e}\}=0,\quad {\{x^{\mu },x^{\nu }\}}%
^{2}=0.
\]
The last equation says that the determinant $g=\det g_{ab}$ of the induced
metric $g_{ab}=\partial _{a}x\cdot \partial _{b}x$ is zero, i.e. $g\equiv 
\frac{1}{2}{\{x^{\mu },x^{\nu }\}}^{2}=0$. The fact that the metric is
degenerate implies that the world-sheet of the string is a null surface.
Also, the tension of the null string is zero\footnote{%
The string with tension 1/$\alpha ^{\prime }$ is obtained by adding a
cosmological constant by analogy to the massive particle action $S=\int
d^{2}\sigma \left[ \frac{\left\{ x^{\mu },x^{\nu }\right\} ^{2}}{2e}{-}\frac{%
e}{\left( 2{\alpha }^{\prime }\right) ^{2}}\right] =\frac{-1}{\alpha
^{\prime }}\int d^{2}\sigma \sqrt{-g}$ as seen by integrating out $e$.
Recall that the massive particle action is obtained as a gauge choice of the
gauge invariant action (\ref{action}, \ref{confor}) and the mass is a
modulus (value of a field \cite{b3}) . We expect the tension to arise in a
similar way in the formalism of Two-Time-Physics. }. We note that a solution
of the equations of motion is given by strings that can be excited purely
for left or purely right movers (keep either $+$ or $-$ ) with the following
constraints 
\begin{eqnarray}
x^{\mu }\left( \tau ,\sigma \right)  &=&q^{\mu }+p^{\mu }\tau -i\sum_{n}%
\frac{1}{n}\alpha _{n}^{\mu }\,e^{in\left( \tau \pm \sigma \right) }, \\
\alpha _{n}^{\mu } &=&\alpha _{n}p^{\mu }\,\,\,\,if\,\,\,\,p^{2}\neq 0,\quad
p\cdot \alpha _{n}=0\,\,\,\,if\,\,\,p^{2}=0,\quad \alpha _{n}^{\mu
}=any\,\,if\,\,p^{\mu }=0.\quad 
\end{eqnarray}

Just like the massless particle action, the null string action is invariant
under the global $d$-dimensional conformal group $SO(d,2)$ for which 
\begin{eqnarray}
\delta x^\mu &=&\lambda x^\mu +a^\mu +\left( \frac 12b^\mu x^2-b\cdot xx^\mu
\right) +\frac 12\omega ^{\mu \nu }x_\nu  \label{confgr} \\
\delta e &=&2\left( \lambda -b\cdot x\right) e  \nonumber
\end{eqnarray}
where $\lambda ,a^\mu ,b^\mu ,\omega ^{\mu \nu }$ are the infinitesimal
global parameters for dilatations, translations, conformal transformations
and Lorentz transformations in Minkowski spacetime. We can reinterpret this $%
SO(d,2)$ conformal symmetry as the two-time Lorentz symmetry if we start
from an action that is manifestly $SO(d,2)$ invariant in $d+2$ dimensions,
but has extra local symmetries, so that the gauge fixed version of such an
action reduces to a $d$-space-time dimensional action for the null string (%
\ref{nulls}). Thus consider the following manifestly $SO(d,2)$ invariant
action as a generalization of the particle action (\ref{confor}) 
\begin{equation}
S_1=\int d^2\sigma \left[ \frac{\left\{ X^M,X^N\right\} ^2}{2e}-\frac
12K\,X\cdot X\right]  \label{str1}
\end{equation}
In addition to $\tau ,\sigma $ reparametrization invariance with parameter $%
\varepsilon ^a\left( \tau ,\sigma \right) $%
\begin{equation}
\delta _\varepsilon X^M=\varepsilon ^a\partial _aX^M,\quad \delta
_\varepsilon e=\partial _a\left( \varepsilon ^ae\right) ,\quad \delta
_\varepsilon K=\partial _a\left( \varepsilon ^aK\right) ,
\end{equation}
the action is also invariant under the local scale transformations with
parameter $\alpha (\tau ,\sigma )$ 
\begin{equation}
\delta _\alpha X^M=\alpha X^M,\quad \delta _\alpha e=4\alpha e,\quad \delta
_\alpha K=-2\alpha K+\left\{ X^M,\frac 1e\left\{ \alpha ,X_M\right\} \right\}
\end{equation}
These generalize the symmetries (\ref{dtau}, \ref{dalpha}) from particles to
strings.

Using the local scale invariance we can choose again the gauge $X^{+^{\prime
}}\left( \tau ,\sigma \right) =1.$ Then the equation of motion for the gauge
potential $K$ (which also acts as a lagrange multiplier for the condition $%
X^2=0$) can be solved for $X^{-^{\prime }}=x^2/2$. Plugging back this
particular gauge choice into the $SO(d,2)$ invariant action (\ref{str1}), we
immediately get the ordinary null string action in d-space-time dimensions (%
\ref{nulls}).

One can also examine the equations of motion for the components $%
X^{+^{\prime }}$, $X^{-^{\prime }}$ and $X^\mu $ to see that they indeed
reduce to the equations of motions of the d-space-time dimensional null
string. Thus, the d-space-time dimensional action (\ref{nulls}) is a genuine
gauge fixed version of the manifestly $SO(d,2)$ invariant action (\ref{str1}%
)!

The generators of the Lorenz transformations in the $(d,2)$- dimensional
space as derived from the action (\ref{str1}) are $L^{MN}=\int_{0}^{2\pi
}d\sigma \,\,(X^{M}P_{0}^{N}$ $-X^{N}P_{0}^{M}$), where now $%
P_{0,1}^{M}=\partial _{0,1}X^{M}$. It is a simple matter to check that by
using the gauge $X^{+^{\prime }}=1$, this reduces to the ordinary generators
of the d-dimensional conformal group (\ref{confgr}) as they would be derived
from the action (\ref{nulls}).

We can immediately extend our discussion to the case of null branes. The
manifestly $SO(d,2)$ invariant action for a null p-brane ($p\geq 2$) is
given by 
\begin{equation}
S_{p}=\int d^{p+1}\sigma \,\,\left[ \frac{%
A^{M_{1}..M_{p+1}}A_{M_{1}...M_{p+1}}}{2e}-\frac{1}{2}KX^{2}\right] 
\end{equation}
where $\det \left( g_{ab}\right) =A^{M_{1}..M_{p+1}}A_{M_{1}...M_{p+1}}$
with the following definition 
\begin{equation}
A^{M_{1}...M_{p+1}}=\epsilon ^{a_{1}...a_{p+1}}\partial
_{a_{1}}X^{M_{1}}...\partial
_{a_{p+1}}X^{M_{p+1}}=\{X^{M_{1}},X^{M_{2}},...,X^{M_{p+1}}\},
\end{equation}
where now $a_{k}$ denote the world-volume indices. As before, by using the
local scale invariance, and going to the gauge $X^{+^{\prime }}=1$, one
derives the standard d-space-time dimensional action for the null p-brane.
The equations of motion of the manifestly $SO(d,2)$ invariant tensionless
p-brane reduce in the same gauge to the equations of motion in d-space-time
dimensions.

Thus we conclude that the physics of both null strings and branes are
invariant under the action of the d-dimensional conformal group $SO(d,2)$
and can be viewed from a ($d,2$)-dimensional point of view.

We have also gained a new perpective for these systems. The action (\ref
{str1}) can now be gauge fixed in many ways as in the particle case \cite
{bda}-\cite{bdm1}$.$ This produces various field theories in a $p+1$
worldvolume with $\left( d-1\right) +1$ degrees of freedom. Since these have
have a common origin, a duality exists between them, which amounts to gauge
transformations involving the local symmetries. We examine one class of such
gauges in the next section.

\section{Tensionless p-Branes in AdS$_{d-n}\times $S$^n$ Backgrounds}

In this section we briefly discuss a class of gauges for the theory defined
by the action (\ref{str1}) in analogy to the particle case in \cite{b3}. We
work in the $d+2$ dimensional coordinate system labelled by $M=\left(
+^{\prime },-^{\prime },\mu ,i\right) $ where the $\left( d-1\right) +1$
dimensions are divided into two sets: the first labelled by $\mu \,$%
describes $\left( d-n-2\right) +1$ dimensions including one time, and the
second labelled by $i$ describes $n+1$ spacelike dimensions. We choose the
gauge that makes the length of the vector $X^i\left( \tau ,\vec{\sigma}%
\right) $ a constant $R$ independent of $\tau ,\vec{\sigma}$, i.e. $\left|
X^i\left( \tau ,\vec{\sigma}\right) \right| =R,$ and then solve the
constraint $X^2=0$. The solution is conveniently parametrized by the two
vectors ${\bf u}^i\left( \tau ,\vec{\sigma}\right) $ and $x^\mu \left( \tau ,%
\vec{\sigma}\right) $%
\begin{equation}
X^i=R\frac{{\bf u}^i}{\left| {\bf u}\right| },\quad X^\mu =\frac{\left| {\bf %
u}\right| }Rx^\mu ,\quad X^{+^{\prime }}=\left| {\bf u}\right| ,\quad
X^{-^{\prime }}=\frac{R^2+{\bf u}^2x^2}{2R^2\left| {\bf u}\right| }.
\end{equation}
The induced metric is 
\begin{eqnarray}
g_{ab} &=&\partial _aX^M\partial _bX^N\eta _{MN} \\
&=&\frac{R^2}{{\bf u}^2}\partial _a{\bf u}^i\partial _b{\bf u}^i+\frac{{\bf u%
}^2}{R^2}\partial _ax^\mu \partial _bx_\mu \\
&=&\frac{R^2}{u^2}\partial _au\partial _bu+\frac{u^2}{R^2}\partial _ax^\mu
\partial _bx_\mu +\partial _a{\bf \Omega }^i\partial _b{\bf \Omega }^i
\end{eqnarray}
where in the last line we have used the definition of the the length of the
vector $u=\left| {\bf u}\right| $ and the unit vector ${\bf \Omega }^i=\frac{%
{\bf u}^i}{\left| {\bf u}\right| }$ in $n+1$ dimensions that describes the
sphere $S^n$. The form of the induced metric $g_{ab}$ shows that the
background metric describes the curved space AdS$_{d-n}\times $S$^n$ given
by 
\begin{equation}
ds^2=\frac{R^2}{u^2}\left( du\right) ^2+\frac{u^2}{R^2}\left( dx^\mu \right)
^2+\left( d{\bf \Omega }\right) ^2.
\end{equation}
Therefore this choice of gauge corresponds to a null p-brane propagating in
an AdS$_{d-n}\times $S$^n$ background described by the lagrangian density 
\begin{equation}
L={\frac 1{2e}}det(g_{ab}).
\end{equation}
This is of interest in the recent literature on the AdS/CFT duality \cite
{m1, w1, gkp}. Our form shows that the global symmetry is larger than the
Killing symmetries of the background $SO\left( d-n-1,2\right) \otimes
SO\left( n+1\right) $; the full symmetry is SO$\left( d,2\right) $ which
contains more than the Killing symmetries.

The important point is that both the flat null $p$-branes of (\ref{nulls})
and the null AdS$\times $S p-branes for every $n,$ appear as particular
gauge choices of the same manifestly $SO(d,2)$ invariant action! This
provides some examples of dual $p$-brane theories. We expect that the
further exploration of other gauges and dualities would be quite interesting.

\section{Higher Order Terms in the Lagrangian: the Rigid String and Branes}

We can also discuss the presence of higher order terms in the manifestly $%
SO(d,2)$ invariant lagrangian we have used in the discussion of the null
string. It turns out that the rigid string \cite{rigid} naturally appears in
this formulation.

First we recall that the action for the rigid string in d-space-time
dimensions is given by \cite{rigid} 
\begin{equation}
S_{rigid} = \int d^{2} \sigma \sqrt{-g}(K_{a}^{ib} K_{b}^{ia})^{2}.
\end{equation}
Here the second fundamental form $K_{ab}^{i}$ is defined by the equation

\begin{equation}
\partial _a\partial _bx^\mu =\Gamma _{ab}^c\partial _cx^\mu +K_{ab}^in_i^\mu
,
\end{equation}
where $\Gamma _{ab}^c=\frac 12g^{cd}(\partial _ag_{db}+\partial
_bg_{da}-\partial _dg_{ab})$ is the Christoffel symbol for the induced
metric $g_{ab}=\partial _ax^\mu \partial _bx^\mu $. Also, the vectors $%
n_i^\mu $ satisfy 
\begin{equation}
n_i^\mu n_j^\mu =\delta _{ij},\quad n_i^\mu \partial _ax^\mu =0,
\end{equation}
where $i=1,...,(d-2)$. The above action can be rewritten in many ways up to
a total derivative \cite{rigid}. For example 
\begin{equation}
S_{rigid}=\int d^2\sigma \sqrt{-g}(\Delta (g)x^\mu )^2,  \label{rig}
\end{equation}
where $\Delta (g)x^\mu =-{\frac 1{\sqrt{-g}}}\partial _a(\sqrt{-g}%
g^{ab}\partial _bx^\mu )$ is the Laplacian defined with respect to the
induced metric $g_{ab}$. Another rewriting of the same action, which we find
particularly useful for our purposes, is \cite{rigid} 
\begin{equation}
S_{rigid}=\int d^2\sigma \sqrt{-g}g^{ab}\partial _at^{\mu \nu }\partial
_bt^{\mu \nu },  \label{rigidd}
\end{equation}
where $t^{\mu \nu }={\frac 1{\sqrt{-g}}}\{x^\mu ,x^\nu \}$.

Now we can easily see that (\ref{rigidd}) is invariant under the action of
the d-dimensional conformal group $SO(d,2)$. We can directly prove the
invariance under dilatations (as originally observed in \cite{rigid}) and
the special conformal transformations (as recently observed in \cite{bcfm})
for the case of the action (\ref{rig}), if we recall the results from
section 3.

Consequently, we can extend the action (\ref{rigidd}) to a space-time with
one extra space and one extra time dimension to the manifestly $SO(d,2)$
invariant action 
\begin{equation}
S_{R1}=\int d^2\sigma \left[ \sqrt{-g}g^{ab}\partial _at^{MN}\partial
_bt^{MN}-\frac 12KX^2\right] ,  \label{R1}
\end{equation}
with $t^{MN}={\frac 1{\sqrt{-g}}}\{X^M,X^N\}$ and $g_{ab}=\partial
_aX^M\partial _bX^N$ $\eta _{MN}$ . In other words, we claim that when (\ref
{R1}) is evaluated in the gauge $X^{+^{\prime }}=1$, $X^{-^{\prime }}=x^2/2$
and $X^\mu =x^\mu $ it reduces to the d-space-time dimensional action (\ref
{rigidd}). The usual conformal transformations are generated again by $%
L^{MN} $ of section 3. Again $L^{MN}$ are naturally interpreted as Lorentz
generators in (d,2) -dimensional space-time.

We can immediately extend this result for the case of rigid branes. The
manifestly $SO(d,2)$ invariant action for a rigid p-brane is 
\begin{equation}
S_{Rp}=\int d^{p+1}\sigma \left[ \sqrt{-g}g^{ab}\partial
_at^{M_1..M_{p+1}}\partial _bt^{M_1...M_{p+1}}-\frac 12KX^2\right] ,
\end{equation}
with 
\begin{eqnarray}
t^{M_1...M_{p+1}} &=&{(-g)}^{\frac{{p+3}}{{4(p+1)}}}\epsilon
^{a_1...a_{p+1}}\partial _{a_1}X^{M_1}...\partial _{a_{p+1}}X^{M_{p+1}} \\
&=&{(-g)}^{\frac{{p+3}}{{4(p+1)}}}\{X^{M_1},X^{M_2},...,X^{M_{p+1}}\}.
\end{eqnarray}
Once again, in the gauge $X^{+^{\prime }}=1$, $X^{-^{\prime }}=x^2/2$ and $%
X^\mu =x^\mu $ this action reduces to the d-space-time dimensional action.

Obviously the same observations can be made if we consider a linear
combination of the manifestly $SO(d,2)$ invariant classical null string (or
brane) and rigid string (or brane) actions in the same gauge.

It is tempting to speculate (as in \cite{bcfm}) that a supersymmetric
version of these tensionless strings is relevant for the understanding of
the dynamics of tensionless strings in 6-space-time dimensions. To make this
speculation more concrete we need to extend our formalism to target-space
supersymmetry, perhaps along the lines of \cite{bdm1, bdm2}.

\section{Outlook}

In this letter we have made a few preliminary steps towards extending the
formalism of Two-Time Physics for the case of string and p-brane ($p\ge 2$)
dynamics. We have found a natural action of the global $SO(d,2)$ symmetry in
the case of several classical systems involving bosonic strings and branes.
The list includes the bosonic null string and branes, the null string moving
in the AdS$\times $S background and the bosonic rigid string and branes.

The next natural question is to ask whether we can extend our results to
other examples of string and brane dynamics. For example, in the case of
particle dynamics it is possible to obtain the massive relativistic particle
as a gauge choice \cite{b3}. Is it analogously possible to relate the
dynamics of strings and branes with tension (see footnote 3) to tensionless
strings and branes such that they are all derived from the same two-time
physics action? One positive indicator in this direction is the observation
made in \cite{town}: the tension of a bosonic string or brane can be
generated if a kind of bosonic Wess-Zumino term is added to the action of
the tensionless string. This mechanism has a natural supersymmetric
extension. It would be interesting to see if this mechanism has a
counterpart in the formalism of Two-Time Physics.

As we have pointed out in the introduction and in eqs.(\ref{action}-\ref
{confor}), one way to think about the gauging of the canonical $Sp(2,R)$
group in the case of particle dynamics, is to identify $Sp(2,R)\sim SO(1,2)$
as the conformal group on the world-line (similarly OSp$\left( n/2\right) $
for conformal supergravity \cite{bd1}). Then various $0+1$ (super) quantum
gravity systems emerge as gauge fixed versions of the same {\it %
(super)conformal} gravity system with the manifest global $SO(d,2)$
symmetry. What is the analog of this viewpoint in the case of strings and
p-branes? It would be natural to extend gravity to conformal gravity on the
world-sheet (or world p+1-volume). For strings the world-sheet conformal
group is $SO(2,2)\sim SO(1,2)_l$ $\times SO(1,2)_r$ $\sim Sp(2,R)_l\times
Sp(2,R)_r$ which is tantalizing. More generally, for a p-brane conformal
gravity would correspond to the gauging of SO$\left( p+1,2\right) $. A
general formulation of various string and p-brane models along these lines,
with the aim of obtaining a general description of Two-Time Physics is
currently under study.

Finally, it is of crucial importance to consider the space-time
supersymmetric extensions of our results, perhaps by generalizing the novel
formalism of \cite{bdm1, bdm2} to strings and branes.

\vskip .5cm

{\bf Acknowledgements:} We wish to thank T. Eguchi, K. Fujikawa, R. Kallosh,
J. Polchinski, N. Sakai, J. Schwarz, and S. K. Yang for interesting comments
and discussions. We would like to thank Tokyo University and the Yukawa
Institute for Theoretical Physics for hospitality during the final stage of
this work.

\end{document}